\documentclass[12pt]{article}%
\usepackage{amsmath}
\usepackage{amsfonts}
\usepackage{amssymb}
\usepackage{graphicx}%
\providecommand{\U}[1]{\protect \rule{.1in}{.1in}}

\newenvironment{proof}[1][Proof]{\noindent \textbf{#1.} }{\  \rule{0.5em}{0.5em}}
\begin{document}

\title{The recollapse problem of closed isotropic models in second order gravity
theory\thanks{Talk given at the Eleventh Marcel Grossmann Meeting on General
Relativity, Berlin 2006}}
\author{John Miritzis\\Department of Marine Sciences, University of the Aegean \\University Hill, Mytilene 81100, Greece\\E-mail: imyr@aegean.gr}
\maketitle

\begin{abstract}
We study the closed universe recollapse conjecture for positively curved
Friedmann-Robertson-Walker (FRW) models in the Jordan frame of the second
order gravity theory. We analyse the late time evolution of the model with the
methods of the dynamical systems. We find that an initially expanding closed
FRW universe, starting close to the Minkowski spacetime, may exhibit
oscillatory behaviour.

\end{abstract}

\section{Introduction}

Since the discovery that general relativity (GR) can been derived from an
action principle, many nonlinear Lagrangians have been used for the
construction of a metric theory of gravity. The reasons for considering higher
order generalisations of the Einstein-Hilbert action are multiple. Firstly,
there is no \textit{a priori }physical reason to restrict the gravitational
Lagrangian to a linear function of the scalar curvature $R.$ Secondly it is
hoped that higher order Lagrangians would create a first approximation to an
as yet unknown theory of quantum gravity. Thirdly one expects that, on
approach to a spacetime singularity, curvature invariants of all orders ought
to play an important dynamical role. Far from the singularity, when higher
order corrections become negligible, one should recover general relativity.

An other important motivation for considering generalised theories of gravity
is the hope that these theories might exhibit better behaviour near
singularities. General relativity leads to singularities in the spacetimes of
all known cosmological models with ordinary matter. Higher order curvature
corrections in the gravitational action may rectify the problem and lead to
cosmological models free from such pathologies, at the cost of diverging from
a FRW behavior at late times \cite{ruzm}. There is a resurgence of interest in
such theories which naturally arise in string-theoretic considerations (cf.
brane models with Gauss-Bonett terms \cite{lno}). These theories are
considered as alternatives to general relativity in an effort to explain the
accelerating expansion of the universe \cite{cdtt} (for a pedagogical review
see \cite{caro}).

This talk is about the late evolution of a closed FRW universe in vacuum in
the context of the quadratic gravity theory. Section 2 is a short exposition
of the field equations obtained from a Lagrangian which is a function of the
Ricci curvature. The conformal equivalence of these theories with the Einstein
equations having a scalar field as a matter source is discussed. In section 3
we investigate the structure of the four-dimensional dynamical system
describing the evolution of a closed FRW universe in quadratic gravity. We
find that periodic solutions may prevent this universe from recollapse.

\section{Jordan versus Einstein frame}

We consider higher order gravity theories (HOG) in vacuum derived from
Lagrangians of the form
\begin{equation}
L=f\left(  R\right)  \sqrt{-g}, \label{hogl}%
\end{equation}
where $f$ is an arbitrary smooth function and $R$ is the Ricci curvature. By
varying $L$ with respect to the metric tensor, we obtain the vacuum
fourth-order field equations
\begin{equation}
f^{\prime}\left(  R\right)  R_{\mu \nu}-\frac{1}{2}f\left(  R\right)  g_{\mu
\nu}-\nabla_{\mu}\nabla_{\nu}f^{\prime}\left(  R\right)  +g_{\mu \nu}\square
f^{\prime}\left(  R\right)  =0, \label{hoge}%
\end{equation}
where $\square=g^{\mu \nu}\nabla_{\mu}\nabla_{\nu}$ and a prime ($^{\prime}$)
denotes differentiation with respect to $R.$ It is well known (see
\cite{baco}), that under the conformal transformation%
\begin{equation}
\widetilde{g}_{\mu \nu}=f^{\prime}\left(  R\right)  g_{\mu \nu}, \label{conf}%
\end{equation}
the field equations reduce to the Einstein field equations with a scalar field
as a matter source, namely%
\[
\widetilde{G}_{\mu \nu}=T_{\mu \nu}\left(  \widetilde{g},\phi \right)  ,
\]
where%
\[
T_{\mu \nu}\left(  \widetilde{g},\phi \right)  =\partial_{\mu}\phi \partial_{\nu
}\phi-\frac{1}{2}\widetilde{g}_{\mu \nu}\left[  \left(  \partial \phi \right)
^{2}-2V\left(  \phi \right)  \right]  ,
\]
and
\begin{equation}
\phi=\sqrt{\frac{3}{2}}\ln f^{\prime}\left(  R\right)  . \label{scfi}%
\end{equation}
Assuming that (\ref{scfi}) can be solved for $R$ to obtain a function
$R\left(  \phi \right)  ,$ the potential of the scalar field is given by
\begin{equation}
V=\frac{1}{2\left(  f^{\prime}\right)  ^{2}}\left(  Rf^{\prime}-f\right)  .
\label{pote}%
\end{equation}

The conformal equivalence between higher order gravity and general relativity
is sometimes expressed formally as \textquotedblleft HOG = GR + scalar field".
However this \textquotedblleft equation\textquotedblright \ is an
oversimplification of the picture. The two frames\footnote{The term frame
denotes the set of dynamical variables used and is not associated to any
coordinate reference frame. In the literature, the original set of variables
is called the \emph{Jordan frame} and the conformally transformed set is
called the \emph{Einstein frame}.} are mathematically equivalent, but
physically they provide different theories. In the Jordan frame, gravity is
described entirely by the metric $g_{\mu \nu}.$ In the Einstein frame, the
scalar field exhibits a \textquotedblleft non-metric\textquotedblright \ aspect
of the gravitational interaction, reflecting the additional degree of freedom
due to the higher order of the field equations in the Jordan frame.

If matter fields, collectively denoted by $\Psi,$ are present in the Jordan
frame, the corresponding field equations become
\begin{equation}
f^{\prime}R_{\mu \nu}-\frac{1}{2}fg_{\mu \nu}-\nabla_{\mu}\nabla_{\nu}f^{\prime
}+g_{\mu \nu}\square f^{\prime}=T_{\mu \nu}\left(  g,\Psi \right)  . \label{hogm}%
\end{equation}
The generalised Bianchi identities imply that
\[
\nabla^{\mu}T_{\mu \nu}\left(  g,\Psi \right)  =0
\]
Therefore, free particles move on geodesics in the Jordan frame. We pass to
the Einstein frame by conformally transforming (\ref{hogm}) to obtain
\begin{equation}
\widetilde{G}_{\mu \nu}=T_{\mu \nu}\left(  \widetilde{g},\phi \right)
+\widetilde{T}_{\mu \nu}\left(  g,\widetilde{g},\Psi \right)  . \label{confm}%
\end{equation}
Now the Bianchi identities imply that
\[
\widetilde{\nabla}^{\mu}\widetilde{T}_{\mu \nu}\left(  g,\widetilde{g}%
,\Psi \right)  \neq0,\  \  \  \  \widetilde{\nabla}^{\mu}T_{\mu \nu}\left(
\widetilde{g},\phi \right)  \neq0.
\]
Therefore, equations (\ref{confm}) are formally the Einstein equations, but
this theory is not physically equivalent to GR. Of course one could firstly
conformally transform the vacuum Lagrangian (\ref{hogl}) into the Einstein
frame to obtain%
\[
\widetilde{L}=\sqrt{-\widetilde{g}}\left(  \widetilde{R}-\widetilde{g}^{\mu
\nu}\partial_{\mu}\phi \partial_{\nu}\phi-2V\left(  \phi \right)  \right)  ,
\]
and then add the matter Lagrangian $L_{m}\left(  \widetilde{g},\Psi \right)  $
with minimal coupling. In that case, the stress-energy tensors of both fields
$\phi$ and $\Psi$ are separately conserved,%
\[
\widetilde{\nabla}^{\mu}T_{\mu \nu}\left(  \widetilde{g},\Psi \right)
=0,\  \  \  \  \widetilde{\nabla}^{\mu}T_{\mu \nu}\left(  \widetilde{g}%
,\phi \right)  =0.
\]
This discussion raises the question: which is the physical metric? There is no
universally acceptable answer to this question in the literature (see
\cite{maso,fgn} for a thorough presentation of different views).

There is another subtle issue, rarely discussed. The conformal transformation
(\ref{conf}) preserves the causal structure of the two frames only if
$f^{\prime}\left(  R\right)  $ is non-negative on the whole spacetime. A
weaker assumption is that $f^{\prime}$ has the same sign on a connected open
subset of the spacetime. However, both assumptions rely on the knowledge of
the function $R,$ and this is possible only after the field equations have
been solved. In any case, the conformal transformation (\ref{conf}) may fail
to be regular at all points of the spacetime.

\section{Cosmological considerations}

Since it is easier to tackle a problem in the context of general relativity
than using the fourth-order equations (\ref{hoge}), the conformal equivalence
theorem allows certain results valid in general relativity to be transferred
in HOG. If a problem can been solved in the Jordan frame it should be
interesting to compare this result with the solution of the same problem
obtained in the Einstein frame.

We investigate \emph{the closed-universe recollapse conjecture} for the
$f\left(  R\right)  =R+\beta R^{2}$ theory. This conjecture states roughly
that a closed universe cannot expand for ever, provided that the matter
content satisfies some energy conditions and has non-negative pressures. For
homogeneous and isotropic spacetimes in the Einstein frame this conjecture was
found to be true for vacuum models \cite{miri1} and for models with a perfect
fluid with equation of state $p=(\gamma-1)\rho,$ $2/3<\gamma<2$ \cite{miri2}.
More precisely it was shown that, \emph{an initially expanding closed FRW
universe, starting close to the Minkowski spacetime cannot avoid recollapse}.
For the $R+\beta R^{2}$ theory in the Einstein frame in vacuum, the state of
the system is defined by $\left(  a,\dot{a},\phi,\dot{\phi}\right)
\in \mathbb{R}^{4}$ and one has a problem of scalar-field cosmology with a
potential (cf. equation (\ref{pote}))
\[
V\left(  \phi \right)  =\frac{1}{8\beta}\left(  1-e^{-\sqrt{2/3}\, \phi
}\right)  ^{2}.
\]

From the general field equations (\ref{hoge}) we obtain the trace equation
\[
\ddot{R}+3H\dot{R}+\frac{1}{6\beta}R=0,
\]
and the $0-0$ equation
\[
\dot{R}=\frac{R^{2}}{12H}-RH-\frac{H}{2\beta}-\frac{Rk}{Ha^{2}}-\frac
{k}{2\beta Ha^{2}}%
\]
in vacuum. Setting $x=1/a$ our dynamical system in vacuum is%
\begin{align}
\dot{R}  &  =v,\nonumber \\
\dot{v}  &  =-3Hv-\frac{1}{6\beta}R,\label{sys1}\\
\dot{x}  &  =-xH,\nonumber \\
\dot{H}  &  =\frac{1}{6}R-2H^{2}-kx^{2},\nonumber
\end{align}
subject to the constraint
\begin{equation}
\left(  1+2\beta R\right)  \left(  H^{2}+kx^{2}\right)  +2\beta Hv=\frac
{\beta}{6}R^{2}. \label{const}%
\end{equation}
Since we are interested only for the closed, $k=+1,$ models, from now on we
omit $k$ from the formulas.

The only equilibrium point is the origin $\left(  0,0,0,0\right)  $.
Linearisation shows that the eigenvalues of the Jacobian matrix at the origin
are $\pm i/\sqrt{6\beta},0,0.$ For nonhyperbolic equilibria the
Hartman-Grobman Theorem does not give any information regarding their
stability and therefore, we shall try to find the normal form of the system.
Let $T$ be the matrix which transforms the linear part of the vector field
into Jordan canonical form. We write (\ref{sys1}) in vector notation as
\begin{equation}
\mathbf{\dot{z}}=A\mathbf{z}+\mathbf{F}\left(  \mathbf{z}\right)  ,
\label{sys3a}%
\end{equation}
where $A$ is the linear part of the vector field and $\mathbf{F}\left(
\mathbf{0}\right)  =\mathbf{0}$. Using the matrix $T$, we define new
variables, $\left(  u,w,y,x\right)  \equiv \mathbf{x}$, by the equations%
\begin{equation}
R=\sqrt{\frac{6}{\beta}}u,\  \ v=-\frac{w}{\beta},\  \ H=w+y,\  \  \ x=x,
\label{litr}%
\end{equation}
or in vector notation $\mathbf{z}=T\mathbf{x},$ so that (\ref{sys3a}) becomes
\[
\mathbf{\dot{x}}=T^{-1}AT\mathbf{x}+T^{-1}\mathbf{F}\left(  T\mathbf{x}%
\right)  .
\]
Denoting the canonical form of $A$ by $B$ we finally obtain the system
\begin{equation}
\mathbf{\dot{x}}=B\mathbf{x}+\mathbf{f}\left(  \mathbf{x}\right)  ,
\label{sys3b}%
\end{equation}
where $\mathbf{f}\left(  \mathbf{x}\right)  :=T^{-1}\mathbf{F}\left(
T\mathbf{x}\right)  .$ In components system (\ref{sys3b}) is%
\begin{align}
\dot{u}  &  =-\frac{1}{\sqrt{6\beta}}w,\nonumber \\
\dot{w}  &  =\frac{1}{\sqrt{6\beta}}u-3w^{2}-3wy,\label{compo}\\
\dot{y}  &  =w^{2}-wy-2y^{2}-x^{2},\nonumber \\
\dot{x}  &  =-wx-yx.\nonumber
\end{align}
Under the non-linear change of variables%
\begin{align}
u  &  \rightarrow u+2\sqrt{6\beta}u^{2}+\sqrt{6\beta}w^{2}+\frac{3}{4}%
\sqrt{6\beta}wy,\nonumber \\
w  &  \rightarrow w+2\sqrt{6\beta}uw+\frac{3}{4}\sqrt{6\beta}uy,\nonumber \\
y  &  \rightarrow y-\frac{\sqrt{6\beta}}{2}uw+\sqrt{6\beta}uy,\label{notr}\\
x  &  \rightarrow x+\sqrt{6\beta}ux,\nonumber
\end{align}
and keeping only terms up to second order, the system (\ref{compo}) transforms
to
\begin{align}
\dot{u}  &  =-\frac{1}{\sqrt{6\beta}}w-\frac{3}{2}yu,\nonumber \\
\dot{w}  &  =\frac{1}{\sqrt{6\beta}}u-\frac{3}{2}yw,\nonumber \\
\dot{y}  &  =\frac{1}{2}\left(  u^{2}+w^{2}\right)  -2y^{2}-x^{2}%
,\label{normal}\\
\dot{x}  &  =-yx.\nonumber
\end{align}

\textbf{Remark}. The results of the normal form theory are valid only near the
origin. Furthermore, the system (\ref{sys1}) is not an arbitrary
\textquotedblleft free\textquotedblright \ four-dimensional system. In view of
the $0-0$ equation, the variables have to satisfy the constraint
(\ref{const}). The nonlinear coordinate transformation (\ref{notr}) is valid
in a neighbourhood of the origin, small enough to ensure that the new
variables still respect the constraint (\ref{const}).

Defining cylindrical coordinates $\left(  u=r\cos \theta,w=r\sin \theta
,y=y,x=x\right)  ,$ we obtain%
\begin{align}
\dot{r}  &  =-\frac{3}{2}ry,\nonumber \\
\dot{\theta}  &  =\frac{1}{\sqrt{6\beta}},\nonumber \\
\dot{y}  &  =\frac{1}{2}r^{2}-2y^{2}-x^{2},\label{cyli}\\
\dot{x}  &  =-yx.\nonumber
\end{align}
We note that the $\theta$ dependence of the vector field has been eliminated,
so that we can study the system in the $(r,y,x)$ space. The second equation of
(\ref{cyli}) implies that the trajectory in the $u-w$ plane spirals with
angular velocity $1/\sqrt{6\beta}.$ We write the first and fourth of
(\ref{cyli}) as a differential equation%
\[
\frac{dr}{dx}=\frac{3}{2}\frac{r}{x},
\]
which has the general solution%
\begin{equation}
r=Ax^{3/2},\  \ A>0. \label{r-equ}%
\end{equation}
We substitute (\ref{r-equ}) into the third equation of (\ref{cyli}) and we
obtain the projection of the forth-dimensional system on the $x-y$ plane,
namely%
\begin{align}
\dot{x}  &  =-yx,\nonumber \\
\dot{y}  &  =bx^{3}-2y^{2}-x^{2},\  \  \ b>0. \label{2dim}%
\end{align}
Some general comments about the trajectories of the system (\ref{2dim}) follow
by inspection. Firstly, (\ref{2dim}) is invariant under the transformation
$t\rightarrow-t,$ $y\rightarrow-y,$ which implies that all trajectories are
symmetric with respect to the $x$ axis. Secondly, the line $x=0$ is invariant
and therefore, a trajectory starting at the half plane $x>0$ remains there for
all $t>0.$ Note that $y$ is decreasing along the orbits in the strip $0<x<1/b$
while $x$ is decreasing in the first quadrant. On any orbit starting in the
first quadrant with $x<1/b,$ $y$ becomes zero at some time and the trajectory
crosses vertically the $x-$axis. Once the trajectory enters the second
quadrant, $x$ increases and $y$ decreases. System (\ref{2dim}) has two
equilibrium points, the origin $\left(  0,0\right)  $ and and $\left(
1/b,0\right)  $ on the $x-$axis. Computation of the Jacobian matrix at the
equilibrium $\left(  1/b,0\right)  $ shows that this point is a center for the
linearised system. Hopefully (\ref{2dim}) has a first integral, \textit{viz}.%
\[
\phi \left(  x,y\right)  =-\frac{2b}{x}+\frac{1}{x^{2}}+\frac{y^{2}}{x^{4}}.
\]
In fact, writing (\ref{2dim}) as
\[
\frac{dy}{dx}=\frac{bx^{3}-2y^{2}-x^{2}}{-yx},
\]
and setting $y^{2}=z,$ we obtain a linear differential equation for $z$ which
is easily integrable. The level curves of $\phi$ are the trajectories of the system.

We shall show for the system (\ref{2dim}) that \emph{(i) there are no solution
curves asymptotically approaching the origin (ii) there exist periodic
solutions and (iii) the basin of attraction of every periodic trajectory is
the set} $y^{2}+x^{2}-2bx^{3}<0.$

\begin{proof}
The function $\phi$ has a local isolated minimum at $\left(  1/b,0\right)  $
and therefore its level curves near this point are closed. For $\phi \left(
x,y\right)  =C$ we have%
\[
y^{2}=x^{2}\left(  -1+2bx+Cx^{2}\right)  ,
\]
which implies that $-1+2bx+Cx^{2}$ must be non-negative. It follows that for
$C>0$ any orbit starting in the first quadrant satisfies%
\[
x\geq \frac{1}{C}\left(  -b+\sqrt{b^{2}+C}\right)  >0,
\]
i.e., there are no solutions approaching $\left(  0,0\right)  .$ For
$C\in \left(  -b^{2},0\right)  $ an orbit of (\ref{2dim}) starting in the first
quadrant crosses the $x-$axis at $\left(  -b-\sqrt{b^{2}+C}\right)  /C$ and
re-enters in the first quadrant crossing the $x-$axis at $\left(
-b+\sqrt{b^{2}+C}\right)  /C,$ i.e. it is a closed curve and represents a
periodic solution. The curve corresponding to $C=0$ separates the phase space
into two disjoint regions I and II. In region I every initially expanding
universe eventually recollapses. In region II, ($C<0$), every trajectory
corresponds to a periodic solution and we conclude that the basin of
attraction of every periodic trajectory is the set $y^{2}+x^{2}-2bx^{3}<0.$
\end{proof}

The phase portrait is shown in Figure 1.

\begin{figure}[h]
\begin{center}
\includegraphics{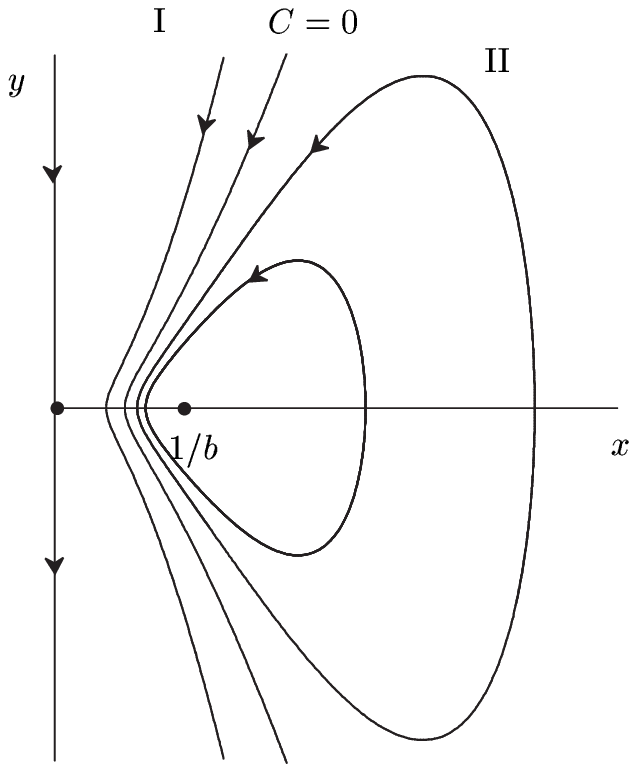}
\end{center}
\caption{Phase portrait of (\ref{2dim})}%
\label{fig1}%
\end{figure}

\section{Discussion}

The periodic solutions of (\ref{2dim}) induce periodicity to the full
four-dimensional system (\ref{cyli}) or (\ref{normal}). In fact, if $x\left(
t\right)  $ and $y\left(  t\right)  $ are periodic solutions, then
(\ref{r-equ}) implies that the functions
\[
u\left(  t\right)  =r\left(  t\right)  \cos \left(  \tfrac{1}{\sqrt{6\beta}%
}t\right)  ,\  \  \ w=r\left(  t\right)  \sin \left(  \tfrac{1}{\sqrt{6\beta}%
}t\right)
\]
oscillate in the $u-w$ plane with a periodic amplitude $r(t).$ Obviously one
cannot assign a physical meaning to the new variables $\left(  u,w,x,y\right)
$ since the transformations (\ref{litr}) and (\ref{notr}) have
\textquotedblleft mixed\textquotedblright \ the original variables of
(\ref{sys1}) in a nontrivial way. However, the periodic character of the
solutions of (\ref{normal}) whatever the physical meaning of the variables be,
has the following interpretation. \emph{Close to the equilibrium of the
original system (\ref{sys1}), there exist periodic solutions for all
variables}. This implies that an initially expanding closed universe can avoid
recollapse through an infinite sequence of successive expansions and
contractions. This interesting result was not revealed in the Einstein frame
\cite{miri2}. Since the basin of attraction of all periodic trajectories of
(\ref{2dim}) is an open subset of the phase space, there is enough room in the
set of initial data of (\ref{sys1}) which lead to an oscillating scale factor.
It should be interesting to investigate the late time evolution of closed FRW
models in the Jordan frame by including matter satisfying some plausible
energy conditions. This issue will be presented elsewhere.

\section*{Acknowledgements}

I thank Spiros Cotsakis and Alan Rendall for useful comments. This work was
co-funded by 75\% from the EU and 25\% from the Greek Government, under the
framework of the \textquotedblleft EPEAEK: Education and initial vocational
training program - Pythagoras\textquotedblright.\

\end{document}